\documentclass[english,aps,superscriptaddress,preprintnumbers,reprint,footinbib,amsmath,amssymb,prb]{revtex4-2}
\usepackage[latin9]{inputenc}
\usepackage{babel}
\usepackage{amsmath}
\usepackage{amssymb}
\usepackage{graphicx}
\usepackage{esint}
\usepackage{hyperref}

\makeatletter

\newcommand{\lyxdot}{.}

\@ifundefined{textcolor}{}{%
 \definecolor{BLACK}{gray}{0}
 \definecolor{WHITE}{gray}{1}
 \definecolor{RED}{rgb}{1,0,0}
 \definecolor{GREEN}{rgb}{0,1,0}
 \definecolor{BLUE}{rgb}{0,0,1}
 \definecolor{CYAN}{cmyk}{1,0,0,0}
 \definecolor{MAGENTA}{cmyk}{0,1,0,0}
 \definecolor{YELLOW}{cmyk}{0,0,1,0}
}

\usepackage{aecompl}

\usepackage{epsfig}\usepackage{dcolumn}\usepackage{bm}
\usepackage{babel}

\hypersetup{
    colorlinks=true,
    linkcolor=blue,
    urlcolor=blue,
    citecolor=blue
        }

\makeatother

\begin{document}
\title{Squeezed state protection of fine structure in ``Poor Man's Majorana''
via quantum spin coupling}
\author{J.E. Sanches}
\email[corresponding author:]{jose.sanches@unesp.br}

\affiliation{São Paulo State University (Unesp), School of Engineering, Department
of Physics and Chemistry, 15385-007, Ilha Solteira-SP, Brazil}
\author{L.T. Lustosa}
\affiliation{São Paulo State University (Unesp), School of Engineering, Department
of Physics and Chemistry, 15385-007, Ilha Solteira-SP, Brazil}
\author{L.S. Ricco}
\affiliation{Science Institute, University of Iceland, Dunhagi-3, IS-107, Reykjavik,
Iceland}
\author{H. Sigurðsson}
\affiliation{Science Institute, University of Iceland, Dunhagi-3, IS-107, Reykjavik,
Iceland}
\affiliation{Institute of Experimental Physics, Faculty of Physics, University
of Warsaw, ulica Pasteura 5, PL-02-093 Warsaw, Poland}
\author{\\M. de Souza}
\affiliation{São Paulo State University (Unesp), IGCE, Department of Physics, 13506-970,
Rio Claro-SP, Brazil}
\author{M.S. Figueira}
\affiliation{Instituto de Física, Universidade Federal Fluminense, 24210-340, Niterói,
Rio de Janeiro, Brazil}
\author{E. Marinho Jr.}
\email[corresponding author:]{enesio.marinho@unesp.br}

\affiliation{São Paulo State University (Unesp), School of Engineering, Department
of Physics and Chemistry, 15385-007, Ilha Solteira-SP, Brazil}
\author{A.C. Seridonio}
\email[corresponding author:]{antonio.seridonio@unesp.br}

\affiliation{São Paulo State University (Unesp), School of Engineering, Department
of Physics and Chemistry, 15385-007, Ilha Solteira-SP, Brazil}
\begin{abstract}
The \textit{``Poor Man's Majorana}'' {[}\href{https://journals.aps.org/prb/abstract/10.1103/PhysRevB.86.134528}{Phys. Rev. B 86, 134528 (2012)}{]}
devoid of topological protection has been theoretically predicted
to rely on the minimal Kitaev chain. Afterward, a pair of superconducting
and \textit{spinless} quantum dots turned the proposal practicable
and differential conductance pinpointed consistent fingerprints with
such a scenario {[}\href{https://www.nature.com/articles/s41586-022-05585-1}{Nature 614, 445 (2023)} and
\href{https://www.nature.com/articles/s41586-024-07434-9}{Nature 630, 329 (2024)}{]}.
In this work, we propose a model wherein the \textit{``Poor Man's
Majorana}'' presents protection when one of the dots is exchange
coupled to a quantum spin. If this quantum dot is perturbed by tuning
the exchange coupling, the well-known spill over-like behavior of
this \textit{Majorana} surprisingly remains unchanged, and solely
half of the fine structure is unexpectedly viewed. As a matter of
fact, the \textit{``Poor Man's Majorana}'' zero mode consists in
squeezing of the other half at zero frequency, which imposes its pinning
there and prevents the mixing of the mode with the explicit fine structure.
We claim that if the supposed unavoidable split of the zero mode by
the fine structure is unexpectedly absent, then the \textit{``Poor
Man's Majorana}'' can be considered robust against the quantum spin.
In this way, it becomes protected and the lack of topological protection
paradigm of the \textit{``Poor Man's Majorana}'' has been revisited,
pushing this seemingly well-established issue into a new direction.
\end{abstract}
\maketitle
\textit{Introduction.-} Proposed by Ettore Majorana in 1937, Majorana
fermions (MFs) consist in real solutions of the Dirac equation in
which the particle is equivalent to its antiparticle{\citep{Majorana-1937}}.
Particularly, in Condensed Matter Physics, such solutions emerge as
quasiparticle excitations known as Majorana bound states (MBSs), which
are zero-energy modes attached to the edges of topological superconductors{\citep{Marra-2022,Leakage-2014,Yuval-Oreg-2022,Oppen-2014,Oppen-2010,Oppen-2013,Klinovaja-2013,Yazdani-2013,Fu-Kane-2009,Fu-Kane-2008,M.-Franz-2013,Nagaosa-2013,Pascal-Simon-2013,Potter-Lee-2012,S-CZhang-2011,Beenakker-2015,Fujimoto-2016,YoichiAndo-2017,Yuval-Oreg-2019,Alicea-2012,Beenakker-2013,Flensberg-2021,Klinovaja-2021,C.Marcus-2016,Mourik-2012,C.Marcus-2017}}.
Within such a context, Alexei Kitaev in 2001, with the so-called Kitaev
toy model{\citep{Kitaev-2001}}, idealized a linear system characterized
by this exotic superconductivity, which is of \textit{p-wave}-type
symmetry, being responsible for the appearance of these non-local
MBSs. Such MBSs, indeed, once showing topological protection, could
be employed as building-blocks for the fault-tolerant quantum computing{\citep{Aguado-2017,Flensberg-2021,Marra-2022}}.
Consequently, the last decade has witnessed a plethora of theoretical
and experimental efforts in exploring potential hosts of MBSs for
technological purposes{\citep{Kitaev-2001,Alicea-2012,Colloquium-Franz-2015,Aguado-2017,Yuval-Oreg-2022,Flensberg-2012,Tewari-2013,Flensberg-2021,Yazdani-2021,Marra-2022}}.

In this route, it is well-kown that, in principle, the realization
of the Kitaev toy model and later on, MBSs at the system boundaries,
depend on the mixing of special ingredients, such as the superconducting
proximity-effect, due to an \textit{s-wave} platform, Zeeman field
and spin-orbit interaction, in particular, over certain systems, such
as linear lattices of magnetic atoms{\citep{Pawlak_2016,J.Franke-2015,Beenakker-2011,Yazdani-2013,Loss-2013,Pascal-Simon-2013,Yazdani-2014,Yazdani-2017,Nitta-2019,Ernst-Meyer-2019,Yazdani-2021,swaveCoupling}}
and semiconducting nanowires{\citep{C.Marcus-2016,Mourik-2012,C.Marcus-2017,Alicea-2012,Colloquium-Franz-2015,Aguado-2017,DasSarma-2010}}.
However, despite these well-established theoretical recipes in bringing
forth MBSs, their detection remains elusive once other phenomena can
misrepresent the MBS\textit{ }signature, such as disorder and Andreev
reflection{\citep{Flensberg-2021,Marra-2022}}. Both of these phenomena
could also yield a zero-energy mode, but topologically trivial instead.

\begin{figure}[!]
\centering\includegraphics[width=1\columnwidth]{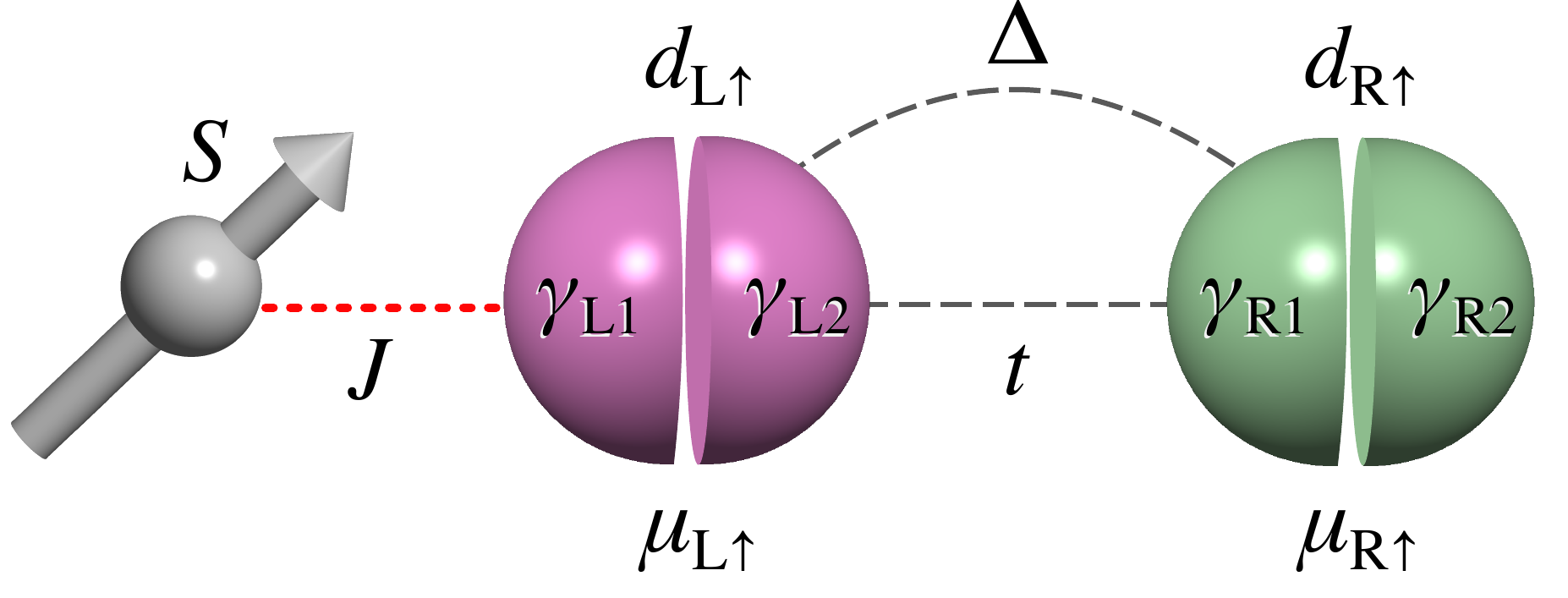} \caption{\label{Fig:Fig.1} Sketch of the minimal Kitaev chain modified by
a quantum spin $S$ exchange coupled via $J$ to the left QD (purple
sphere) with chemical potential $\mu_{L\uparrow}$ in the electronic
basis $d_{L\uparrow},$ which is built-up by the couple of MBSs $\gamma_{L1}$
and $\gamma_{L2}.$ For the right QD (green sphere), we have similarly
$\mu_{L\uparrow},$ $d_{R\uparrow},$ $\gamma_{R1}$ and $\gamma_{R2}.$
Both the QDs are connected to each other by ECT and CAR given by $t$
and $\Delta$ terms, respectively. For $\Delta=t,\mu_{L\uparrow}=\mu_{R\uparrow}=0$
and $J\protect\neq0,$ the zero-energy mode of the \textit{\textquotedblleft Poor
Man's Majorana\textquotedblright{} }$\gamma_{L1}$ is due to the squeezing
of half of the fine structure induced by the quantum spin. As the
other half does not couple to this mode, it becomes protected.}
\end{figure}

Alternatively, the recent experiments reported by Refs.\cite{Kouwenhoven2023,MinimalModel2,MinimalModel3}
with quantum dots (QDs) have pointed out compatible fingerprints with
the so-called \textit{``Poor Man's Majorana}'' (\textit{``Poor
M.M.}'')\cite{Flensberg_2012(Poor),Poor2,Poor3,Poor4,Jelena1,Jelena2}.
Such a designation was introduced by M. Leijnse and K. Flensberg to
label a MBS without topological protection\cite{Flensberg_2012(Poor)}.
They conjectured that in the minimal Kitaev chain, namely that one
with two sites, the MBS with the nickname \textit{``Poor M.M.}''
could be feasible. Such quasiparticle could emerge due to the mixing
of two spin-polarized grounded QDs equally coupled by electronic co-tunneling
(ECT) and crossed Andreev reflection (CAR) processes. In this \textit{sweet
spot}\cite{Flensberg_2012(Poor)}, the isolation of the MBSs\textit{
}would occur in spatially distinct QDs of this dimer. Unfortunately,
due to the lack of topological protection of these MBSs if one of
the QDs is found not grounded, the MBS initially isolated at one QD
then spills over the other.\cite{Flensberg_2012(Poor)}.

In this work, we reveal that the \textit{``Poor M.M.}'' exhibits
a distinct protection if one QD is exchange coupled to a quantum spin.
By perturbing this QD via variations of the exchange coupling, the
well-established spill over-like\textit{ }behavior of the\textit{
``Poor M.M.}''\textit{ }counterintuitively remains, in particular
with the mode pinned at zero frequency and fully decoupled from the
fine structure due to the quantum spin, which is half of the entire
spectrum. Surprisingly, the other half of the fine structure is found
squeezed as the zero-energy mode of the \textit{``Poor M.M.}'',
thus imposing the pinning of such a mode at zero frequency and then,
protecting it against the other half.

Our result makes explicit if the \textit{``Poor M.M.}'' is still
observed arising from this spill over-like\textit{ }behavior, the
protection against a quantum spin is a concrete feature. This is due
to the half of the fine structure clearly detached from this\textit{
}MBS, while the other half squeezed at zero frequency leads to the
formation of the \textit{``Poor M.M.}'' itself. These findings challenge
the lack of topological protection paradigm of the \textit{``Poor
M.M.}'' and provide a new direction for research on this matter.

\textit{The Model.-} We consider the system schematically shown by
Fig.\ref{Fig:Fig.1} and inspired in the experiments reported by Refs.\cite{Kouwenhoven2023,MinimalModel2}.
Distinctly, we account for the fine structure due to a quantum spin
$S$. The simplest way to introduce such is via the \textit{Ising-like}
Hamiltonian $JS^{z}s^{z}$\cite{Ising}. To that end, we adopt the
exchange term $J,$ $S^{z}=\sum_{m}m|m\rangle\langle m|$ wherein
$m=[-S,-S+1,...,S-1,S]$\textit{ }and\textit{ }for the left QD\textit{
}the\textit{ spin-z} operator $s^{z}=\frac{1}{2}\sum_{\sigma}\sigma d_{L\sigma}^{\dagger}d_{L\sigma},$
with $d_{L\sigma}^{\dagger}(d_{L\sigma})$ as the creation (annihilation)
operator and $\sigma=\pm1(\uparrow,\downarrow)$. For the right QD,
we have $d_{R\sigma}^{\dagger}(d_{R\sigma}).$

The QDs are found within the \textit{spinless} regime, where we arbitrary
choose the spin-up channel $\sigma=\uparrow$ as relevant, due to
an imposed large Zeeman splitting. Most relevant, such a dimer of
QDs then constitutes the minimal Kitaev chain, being the QDs linked
to each other by means of ECT and CAR mechanisms, ruled by the hopping
$t$ and superconducting pairing $\Delta$ terms, respectively. This
scenario is mimicked by the effective Hamiltonian
\begin{eqnarray}
{\cal {H}} & = & (\mu_{L\uparrow}+\frac{J}{2}S^{z})d_{L\uparrow}^{\dagger}d_{L\uparrow}+\mu_{R\uparrow}d_{R\uparrow}^{\dagger}d_{R\uparrow}+(td_{L\uparrow}d_{R\uparrow}^{\dagger}\nonumber \\
 & + & \Delta d_{L\uparrow}d_{R\uparrow}+\text{{H.c.}}),\label{eq:H1}
\end{eqnarray}
where $\mu_{L\uparrow(R\uparrow)}$ represents the chemical potential
for the QD $\alpha=L,R.$ Both the electronic operators of the chain
can be projected onto the\textit{ }MBS basis $\gamma_{L1(L2)}$ and
$\gamma_{R1(R2)}$ for the left and the right QDs, respectively. To
perform such, it is imperative to evoke the relations $d_{L\uparrow}=(\gamma_{L1}+i\gamma_{L2})/\sqrt{2}$
and $d_{R\uparrow}=(\gamma_{R1}+i\gamma_{R2})/\sqrt{2}.$ By considering
the \textit{``Majorana chain regime''} $t=\Delta$ in Eq.(\ref{eq:H1})
the Hamiltonian turns into
\begin{equation}
{\cal {H}}=i(\mu_{L\uparrow}+JS^{z})\gamma_{L1}\gamma_{L2}+i\mu_{R\uparrow}\gamma_{R1}\gamma_{R2}+i2t\gamma_{L2}\gamma_{R1},\label{eq:MajoranaChain}
\end{equation}
from where it is notorious the \textit{sweet spot}\cite{Flensberg_2012(Poor)}
$\mu_{\alpha\uparrow}=\mu_{\bar{\alpha}\uparrow}=J=0,$ being $\bar{\alpha}=L(R)$
for the opposite QD $\alpha=R(L),$ characterized by the spatially
apart isolated MBSs $\gamma_{L1}$ and $\gamma_{R2}$ at the left
and right QDs, respectively, once such quasiparticles do not enter
into the Hamiltonian. As we know, these MBSs are the so-called \textit{``Poor
M. Ms.}''\cite{Flensberg_2012(Poor),Kouwenhoven2023,MinimalModel2}.
To understand such, for instance, suppose that $\mu_{R\uparrow}$
is placed off the \textit{sweet spot}. This allows the spectral amplitude
of the zero-energy mode of $\gamma_{R2}$\textit{ }from the right
QD to spill over the left one, in particular on $\gamma_{L2},$ in
such a way that can be visualized together with $\gamma_{L1}.$ As
part of $\gamma_{R2}$ is found at the left QD, the nickname for the\textit{
}MBS $\gamma_{R2}$ is \textit{``Poor M.M.}'', once the topological
protection is lacking.

We reveal by considering $\mu_{\alpha\uparrow}=\mu_{\bar{\alpha}\uparrow}=0$
and $J\neq0$ in the\textit{ ``Majorana chain regime''} that the
zero-energy mode of the \textit{``Poor M.M.}''\textit{ }consists
the feature under protection against the $J$ coupling. It is well-known
that, due to a quantum spin $S,$ the lifting of the energy spectrum
per level into a $2S+1$ fine structure is expected. The zero-energy
mode of the\textit{ ``Poor M.M.}'' then persists even when deviating
from the \textit{sweet spot}, in particular by $J$ and the fine structure
is reduced by half. Equivalently, we perform the break down of the
\textit{sweet spot}, but the zero-energy mode of the \textit{``Poor
M.M.}'' counterintuitively remains pinned and does not belong to
the renormalized fine structure. Indeed, the other half of the fine
structure squeezes as the zero-energy mode of the \textit{``Poor
M.M.}'' and ensures its pinning at $\omega=0,$ thus forbidding the
mixing with the explicit part of the spectrum.

In order to uncover the physical mechanisms within the \textit{``Poor
M.M. regime''} the evaluation of frequency dependent retarded Green's
functions (GFs) for the QDs $\alpha$ are timely, once they dictate
the differential conductance\cite{Kouwenhoven2023,MinimalModel2}.
To this end, we should consider the ordinary spectral densities ${\cal {A}}_{d_{\alpha\uparrow}d_{\alpha\uparrow}^{\dagger}}(\omega)=(-1/\pi)\text{Im}\langle\langle d_{\alpha\uparrow};d_{\alpha\uparrow}^{\dagger}\rangle\rangle$
and ${\cal {A}}_{d_{\alpha\uparrow}^{\dagger}d_{\alpha\uparrow}}(\omega)=(-1/\pi)\text{Im}\langle\langle d_{\alpha\uparrow}^{\dagger};d_{\alpha\uparrow}\rangle\rangle$,
where $\langle\langle A;B\rangle\rangle$ stands for the corresponding
GF. The anomalous GFs ${\cal {A}}_{d_{\alpha\uparrow}^{\dagger}d_{\alpha\uparrow}^{\dagger}}(\omega)=(-1/\pi)\text{Im}\langle\langle d_{\alpha\uparrow}^{\dagger};d_{\alpha\uparrow}^{\dagger}\rangle\rangle$
and ${\cal {A}}_{d_{\alpha\uparrow}d_{\alpha\uparrow}}(\omega)=(-1/\pi)\text{Im}\langle\langle d_{\alpha\uparrow};d_{\alpha\uparrow}\rangle\rangle$
should be taken into account too. As shown below, they determine the
MBS component ${\cal {A}}_{\gamma_{\alpha j}}(\omega)=(-1/\pi)\text{Im}\langle\langle\gamma_{\alpha j};\gamma_{\alpha j}\rangle\rangle$
of the QD. With it, we are able to quantify clearly the MBS from the
QD $\alpha$ that spills over the opposite QD $\bar{\alpha}$ when
the system is driven off the \textit{sweet spot}\cite{Flensberg_2012(Poor),Kouwenhoven2023,MinimalModel2}.
From $\gamma_{L1(L2)}$ and $\gamma_{R1(R2)},$ we obtain the GF as
follows
\begin{eqnarray}
\langle\langle\gamma_{\alpha j};\gamma_{\alpha j}\rangle\rangle & = & \frac{1}{2}[\langle\langle d_{\alpha\uparrow};d_{\alpha\uparrow}^{\dagger}\rangle\rangle+\langle\langle d_{\alpha\uparrow}^{\dagger};d_{\alpha\uparrow}\rangle\rangle\nonumber \\
 & + & \epsilon_{j}(\langle\langle d_{\alpha\uparrow}^{\dagger};d_{\alpha\uparrow}^{\dagger}\rangle\rangle+\langle\langle d_{\alpha\uparrow};d_{\alpha\uparrow}\rangle\rangle)],\label{eq:Expansion}
\end{eqnarray}
where $\epsilon_{j}=+1,-1$ for $j=1,2.$ Thus, the task of calculating
the GFs of Eq.(\ref{eq:Expansion}) can be achieved via the standard
equation-of-motion (EOM) approach\cite{Flensberg-book}, which is
summarized as
\begin{eqnarray}
(\omega+i\Gamma)\langle\langle A;B\rangle\rangle=\langle[A;B]_{+}\rangle+\langle\langle[A,{\cal {\cal {H}}}];B\rangle\rangle,\label{eq:EOM}
\end{eqnarray}
where $\Gamma$ mimics the natural broadening, supposed to be symmetric
for simplicity, arising from the outside environment. By applying
the EOM technique to Eq.(\ref{eq:H1}) for $t\neq\Delta$ we then
find
\begin{eqnarray}
\langle\langle d_{\alpha\uparrow};d_{\alpha\uparrow}^{\dagger}\rangle\rangle & = & \frac{1}{2S+1}\sum_{m}\frac{1}{\omega+i\Gamma-\mu_{\alpha\uparrow}-\frac{Jm}{2}\delta_{\alpha L}-\Sigma_{\alpha}^{+}},\nonumber \\
\label{eq:GF1}
\end{eqnarray}
\begin{eqnarray}
\langle\langle d_{\alpha\uparrow}^{\dagger};d_{\alpha\uparrow}\rangle\rangle & = & \frac{1}{2S+1}\sum_{m}\frac{1}{\omega+i\Gamma+\mu_{\alpha\uparrow}+\frac{Jm}{2}\delta_{\alpha L}-\Sigma_{\alpha}^{-}},\nonumber \\
\label{eq:GF2}
\end{eqnarray}
\begin{eqnarray}
\langle\langle d_{\alpha\uparrow}^{\dagger};d_{\alpha\uparrow}^{\dagger}\rangle\rangle & = & \eta_{\alpha}\frac{1}{2S+1}\sum_{m}\frac{2t\Delta K_{\alpha}^{-}}{\omega+i\Gamma+\mu_{\alpha\uparrow}+\frac{Jm}{2}\delta_{\alpha L}-\Sigma_{\alpha}^{-}},\nonumber \\
\label{eq:GF3}
\end{eqnarray}
and
\begin{eqnarray}
\langle\langle d_{\alpha\uparrow};d_{\alpha\uparrow}\rangle\rangle & = & \eta_{\alpha}\frac{1}{2S+1}\sum_{m}\frac{2t\Delta K_{\alpha}^{+}}{\omega+i\Gamma-\mu_{\alpha\uparrow}-\frac{Jm}{2}\delta_{\alpha L}-\Sigma_{\alpha}^{+}},\nonumber \\
\label{eq:GF4}
\end{eqnarray}
where we used $\langle\langle A;B\rangle\rangle=\sum_{m}\langle\langle A\left|m\right\rangle \left\langle m\right|;B\rangle\rangle,$
the thermal average $\left\langle \left|m\right\rangle \left\langle m\right|\right\rangle =\frac{1}{2S+1}$,
$\delta_{\alpha L}$ as the Kronecker Delta and $\eta_{\alpha}=-1,+1$
for $\alpha=L,R,$ respectively. The self-energy correction due to
the several couplings is $\Sigma_{\alpha}^{\pm}=\tilde{K}_{\bar{\alpha}}^{\pm}+(2t\Delta)^{2}K_{\bar{\alpha}}K_{\alpha}^{\pm},$
with
\begin{eqnarray}
\tilde{K}_{\alpha}^{\pm}=\frac{(\omega+i\Gamma)(t^{2}+\Delta^{2})\pm(\mu_{\alpha\uparrow}+\frac{Jm}{2}\delta_{\alpha L})(t^{2}-\Delta^{2})}{(\omega+i\Gamma)^{2}-(\mu_{\alpha\uparrow}+\frac{Jm}{2}\delta_{\alpha L})^{2}},\nonumber \\
\label{eq:SE1}
\end{eqnarray}
\begin{equation}
K_{\alpha}=\frac{\omega+i\Gamma}{(\omega+i\Gamma)^{2}-(\mu_{\alpha\uparrow}+\frac{Jm}{2}\delta_{\alpha L})^{2}}\label{eq:SE2}
\end{equation}
and
\begin{equation}
K_{\alpha}^{\pm}=\frac{K_{\bar{\alpha}}}{\omega+i\Gamma\pm\mu_{\alpha\uparrow}\pm\frac{Jm}{2}\delta_{\alpha L}-\tilde{K}_{\bar{\alpha}}^{\mp}}.\label{eq:SE3}
\end{equation}

\begin{figure}[!]
\centering\includegraphics[width=1\columnwidth]{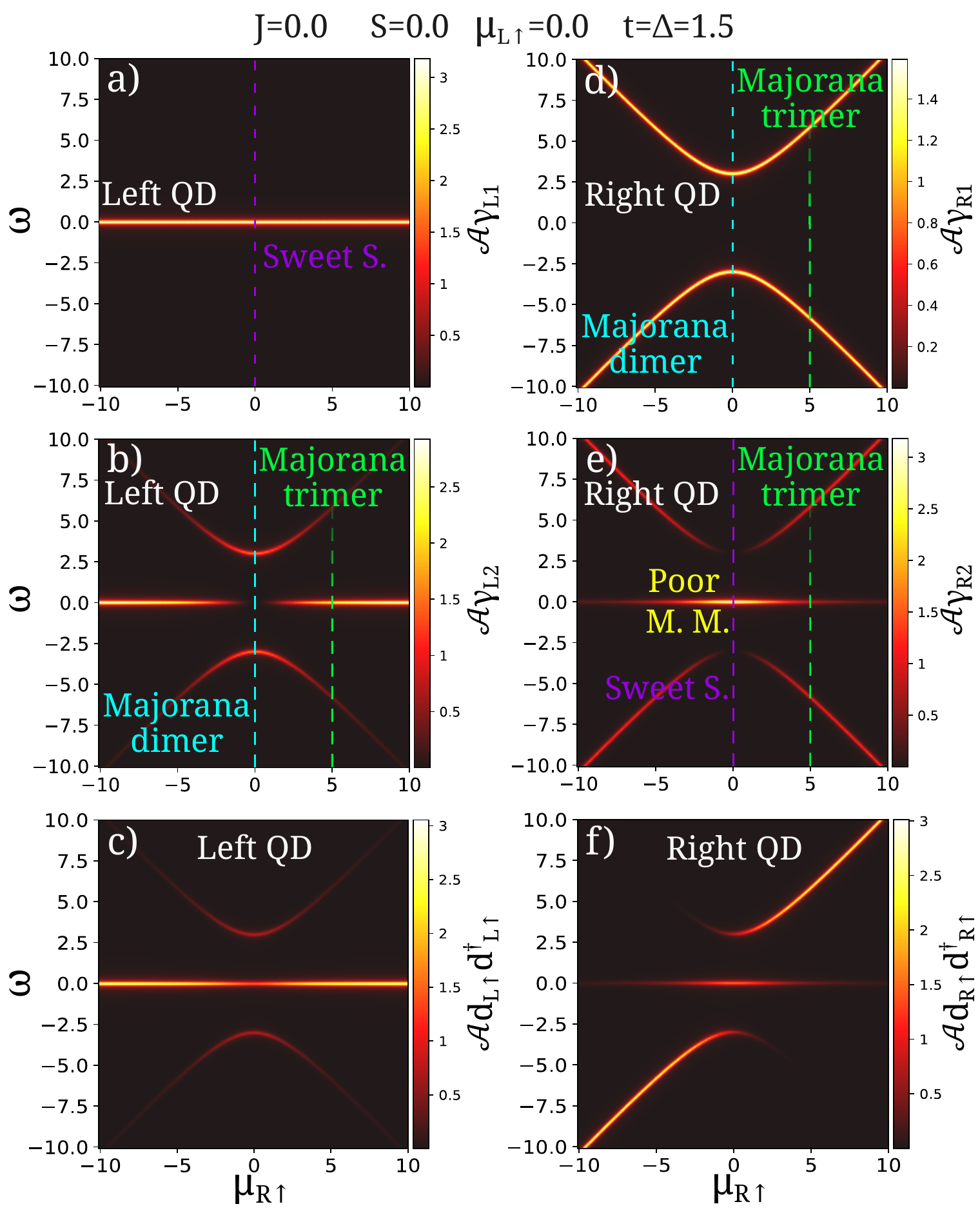} \caption{\label{Fig:Fig.2} Color maps of ${\cal {A}}_{d_{\alpha\uparrow}d_{\alpha\uparrow}^{\dagger}}$
and ${\cal {A}}_{\gamma_{\alpha j}}$ in the \textit{\textquotedblleft Majorana
chain regime\textquotedblright} spanned by $\omega$ and $\mu_{R\uparrow}$.
Panel (a) exhibits a resonant zero-energy mode at $\omega=0$ in ${\cal {A}}_{\gamma_{L1}}$
upon changing $\mu_{R\uparrow}.$ In panel (b) for ${\cal {A}}_{\gamma_{L2}}$
this mode is absent when $\mu_{R\uparrow}=0.$ Upper and lower arcs
emerge. Panel (c) presents ${\cal {A}}_{d_{L\uparrow}d_{L\uparrow}^{\dagger}}$
made by the MBSs $\gamma_{L1}$ and $\gamma_{L2}.$ In panel (d) for
${\cal {A}}_{\gamma_{R1}}$ we have only the upper and lower arcs.
In case of panel (e) for ${\cal {A}}_{\gamma_{R2}}$ there is the
mode at $\omega=0$ and arcs appear off it. As the amplitude of ${\cal {A}}_{\gamma_{R2}}$
at $\omega=0$ decreases while the corresponding of ${\cal {A}}_{\gamma_{L2}}$
increases away from $\mu_{R\uparrow}=0,$ we conclude that $\gamma_{R2}$\textit{
}spills over $\gamma_{L2}$. The left QD is trivial with two locally
zero-energy modes with $\gamma_{L1}$ and $\gamma_{L2}$ contributing
simultaneously. The MBS\textit{ }$\gamma_{R2}$ is then the \textit{\textquotedblleft Poor
Man's Majorana\textquotedblright{} }or simply \textit{\textquotedblleft}Poor
M.M\textit{.\textquotedblright}. On the other hand, the \textit{sweet
spot} (Sweet S.) occurs for fixed $\mu_{R\uparrow}=0,$ where we clearly
see resonant zero-energy modes spatially apart at the left and right
QDs, respectively. This appears via finite values at $\omega=0$ solely
in ${\cal {A}}_{\gamma_{L1}}$ and ${\cal {A}}_{\gamma_{R2}},$ in
opposite to the corresponding null in ${\cal {A}}_{\gamma_{L2}}$
and ${\cal {A}}_{\gamma_{R1}},$ respectively. Panel (f) shows ${\cal {A}}_{d_{R\uparrow}d_{R\uparrow}^{\dagger}}$
made by $\gamma_{R1}$ and $\gamma_{R2}.$ We call attention that
such a scenario breaks down upon varying $\mu_{R\uparrow}$ and thus
leads to the \textit{\textquotedblleft}Poor M.M\textit{.\textquotedblright .}}
\end{figure}

\begin{figure}[!]
\centering\includegraphics[width=1\columnwidth]{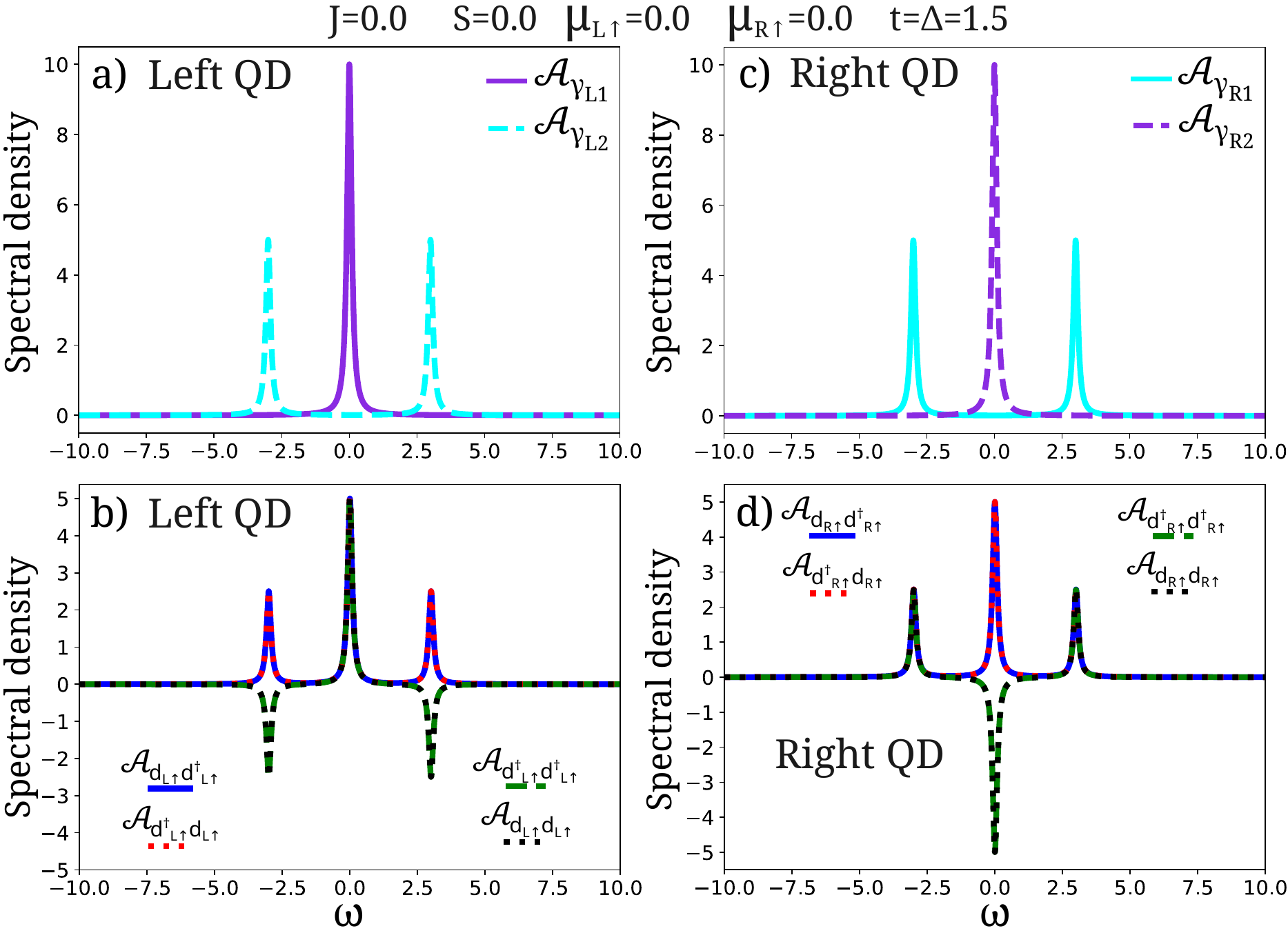} \caption{\label{Fig:Fig.3} Spectral densities in the \textit{\textquotedblleft Majorana
chain regime\textquotedblright{} }within the\textit{ sweet spot.} Panels
(a) and (c) exhibit resonant zero-energy modes spatially apart at
the left and right QDs, respectively. Such a feature can be seen in
${\cal {A}}_{\gamma_{L1}}$ and ${\cal {A}}_{\gamma_{R2}}$ at $\omega=0,$
while ${\cal {A}}_{\gamma_{L2}}$ and ${\cal {A}}_{\gamma_{R1}}$
show a split-peak structure due to the coupling $i2t\gamma_{L2}\gamma_{R1}$
in Eq.(\ref{eq:MajoranaChain}). In panels (b) and (d) the ordinary
spectral densities ${\cal {A}}_{d_{\alpha\uparrow}d_{\alpha\uparrow}^{\dagger}}$
and ${\cal {A}}_{d_{\alpha\uparrow}^{\dagger}d_{\alpha\uparrow}}$
reflect these two resonant zero-energy modes, as well as the satellite
peaks. Notice that the corresponding anomalous ${\cal {A}}_{d_{\alpha\uparrow}^{\dagger}d_{\alpha\uparrow}^{\dagger}}$
and ${\cal {A}}_{d_{\alpha\uparrow}d_{\alpha\uparrow}}$ are phase
shifted by $\pi$ by swapping sides.}
\end{figure}

\begin{figure}[!]
\centering\includegraphics[width=1\columnwidth]{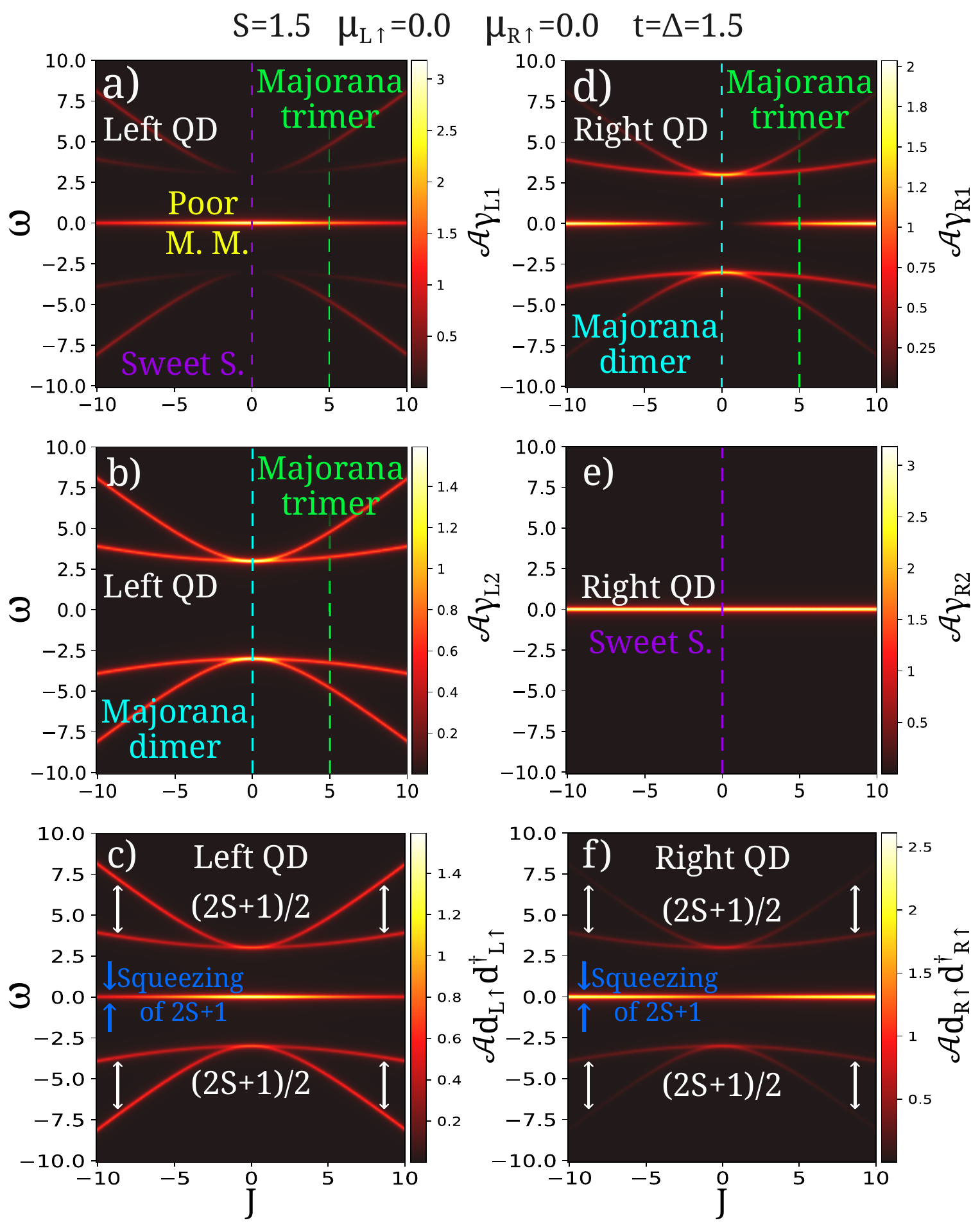} \caption{\label{Fig:Fig.4} Color maps of ${\cal {A}}_{d_{\alpha\uparrow}d_{\alpha\uparrow}^{\dagger}}$
and ${\cal {A}}_{\gamma_{\alpha j}}$ in the \textit{\textquotedblleft Majorana
chain regime\textquotedblright} spanned by $\omega$ and $J$ showing
our main finding: in the presence of the quantum spin $S$, half of
the fine structure is made explicit. The\textit{ }spill over-like
behavior of the zero-energy mode of the \textit{\textquotedblleft Poor
Man's Majorana}\textquotedblright{} continues and the mode is protected
against $J.$ Such a phenomenon is due to the other half that squeezes
itself as the \textit{\textquotedblleft Poor Man's Majorana}\textquotedblright{}
at $\omega=0$ and avoids the mixing with the latter.}
\end{figure}

\textit{Results.-} Throughout the analysis we set $\mu_{L\uparrow}=0$
and the \textit{``Majorana chain regime''} $t=\Delta=1.5$ in arbitrary
units for Figs.\ref{Fig:Fig.2}, \ref{Fig:Fig.3} and \ref{Fig:Fig.4}.

Fig.\ref{Fig:Fig.2} shows the spectral densities spanned by the frequency
$\omega$ and $\mu_{R\uparrow},$ with $\mu_{L\uparrow}=J=0.$ Panel
(a) of Fig.\ref{Fig:Fig.2} exhibits ${\cal {A}}_{\gamma_{L1}}$ for
the left QD, where we see a resonant profile pinned at $\omega=0$
as a zero-energy mode upon tuning $\mu_{R\uparrow}.$ It represents
the isolated MBS $\gamma_{L1}$ of the left QD (purple line cut),
which is found decoupled from any MBS of the system due to the characteristic
$\mu_{L\uparrow}=0$ in Eq.(\ref{eq:MajoranaChain}). At the \textit{sweet
spot} with $\mu_{R\uparrow}=0$ (purple line cuts) solely $\gamma_{L2}$
and $\gamma_{R1}$ couple to each other and build a MF dimer. This
consists of a type of molecule without the mode $\omega=0$ and with
the split levels bounding and anti-bounding states represented by
bottom and top arcs in both the panels (b) and (d) of Fig.\ref{Fig:Fig.2}.
In Fig.\ref{Fig:Fig.2}, see the cyan line cuts as the example of
the MF dimer. The opposite situation is found in ${\cal {A}}_{\gamma_{R2}}$
of Fig. \ref{Fig:Fig.2}(e), which identifies the isolated MBS $\gamma_{R2}$
(purple line cut). Upon varying $\mu_{R\uparrow},$ the MF trimer
composed by $\gamma_{L2},\gamma_{R1}$ and $\gamma_{R2}$ is established
instead. The trimer formation is symbolized by panels (b), (d) and
(e) of Fig.\ref{Fig:Fig.2}, where the bonding (bottom arc) and anti-bonding
(top arc) states appear together with the mode at $\omega=0,$ the
so-called non-bonding, also in Figs.\ref{Fig:Fig.2}(b) and (e). To
note this, see in Fig.\ref{Fig:Fig.2} the green line cuts as the
example of the MF trimer. In this manner, the \textit{``Poor M.M.
regime''} emerges, being characterized by the amplitude of ${\cal {A}}_{\gamma_{R2}}$
at $\omega=0$ that decreases while the corresponding in ${\cal {A}}_{\gamma_{L2}}$
increases by changing $\mu_{R\uparrow}.$ It means that the initially
isolated MBS $\gamma_{R2}$ at the right QD spills over the left QD
and becomes the \textit{``Poor M.M.}''\cite{Flensberg_2012(Poor)}.
Such a behavior is also recorded in ${\cal {A}}_{d_{L\uparrow}d_{L\uparrow}^{\dagger}}$
and ${\cal {A}}_{d_{R\uparrow}d_{R\uparrow}^{\dagger}},$ which at
the \textit{sweet spot} with $\mu_{R\uparrow}=0$ compete at equal
footing, i.e, ${\cal {A}}_{d_{L\uparrow}d_{L\uparrow}^{\dagger}}(0)={\cal {A}}_{d_{R\uparrow}d_{R\uparrow}^{\dagger}}(0).$
However, the \textit{``Poor M.M. regime''} $\mu_{R\uparrow}\neq0,$
the unbalance ${\cal {A}}_{d_{R\uparrow}d_{R\uparrow}^{\dagger}}(0)<{\cal {A}}_{d_{L\uparrow}d_{L\uparrow}^{\dagger}}(0)$
takes place due to the spill over-like behavior from $\gamma_{R2}$
to $\gamma_{L2}.$

Fig.\ref{Fig:Fig.3} makes explicit the \textit{sweet spot}. Particularly,
Figs.\ref{Fig:Fig.3}(a) and (c) show the features of $\gamma_{L1}$
and $\gamma_{R2}$ that clearly reveal the resonant states at $\omega=0$,
thus corresponding to the purple line cuts marked in Fig.\ref{Fig:Fig.2}(a)
and (e), respectively. On the other hand, $\gamma_{L2}$ also presented
in Fig.\ref{Fig:Fig.3}(a), exhibits two resonant states around $\omega=0$
instead, due to the finite coupling $t=\Delta$ of the MF dimer made
by $\gamma_{L2}$ and $\gamma_{R1}.$ As aftermath, the same behavior
of $\gamma_{L2}$ is followed by $\gamma_{R1},$ as can be verified
in Fig.\ref{Fig:Fig.3}(c). It means that two isolated MBSs $\gamma_{L1}$
and $\gamma_{R2}$ are spatially apart at the left and right QDs,
respectively. For completeness, in Figs.\ref{Fig:Fig.3}(b) and (d)
we summarize the spectral profiles of the GFs ${\cal {A}}_{d_{\alpha\uparrow}d_{\alpha\uparrow}^{\dagger}},{\cal {A}}_{d_{\alpha\uparrow}^{\dagger}d_{\alpha\uparrow}},{\cal {A}}_{d_{\alpha\uparrow}^{\dagger}d_{\alpha\uparrow}^{\dagger}}$
and ${\cal {A}}_{d_{\alpha\uparrow}d_{\alpha\uparrow}}$ of the QDs.
Interestingly enough, the anomalous GFs, namely ${\cal {A}}_{d_{\alpha\uparrow}^{\dagger}d_{\alpha\uparrow}^{\dagger}}$
and ${\cal {A}}_{d_{\alpha\uparrow}d_{\alpha\uparrow}},$ exhibit
spectral profiles shifted by $\pi$ by making the swap $L\leftrightarrow R,$
once these GFs depend on the parity of the QD site.

Now in Fig.\ref{Fig:Fig.4} we discuss our main finding by introducing
the \textit{protection-like} behavior observed in the zero-energy
mode arising from the \textit{``Poor M.M.''}. For such an analysis,
we consider $S=1.5$, $\mu_{L\uparrow}=\mu_{R\uparrow}=0$ and evaluate
the spectral densities spanned by $\omega$ and $J.$ We call attention
that, according to Eq.(\ref{eq:MajoranaChain}), the exchange $J$
plays the role of an effective chemical potential acting over the
left QD. It means that when $J=0$ the \textit{sweet spot }is restored
and we have again the MF dimer $\gamma_{L2}$ and $\gamma_{R1},$
with the non-local and isolated $\gamma_{L1}$ and $\gamma_{R2}$
at the left and right QDs, respectively. However for $J\neq0,$ the\textit{
}MF trimer becomes now composed by $\gamma_{L1},\gamma_{L2}$ and
$\gamma_{R1},$ thus driving the system into the \textit{``Poor M.M.
regime''}, where the novel \textit{``Poor M.M.'' }is expected to
be $\gamma_{L1}.$

We begin the numerical analysis with ${\cal {A}}_{\gamma_{L1}}$ and
${\cal {A}}_{\gamma_{R2}}$ in Figs.\ref{Fig:Fig.4}(a) and (e), which
due to the $J=0$ condition, gives rise to the non-local and isolated
MBSs $\gamma_{L1}$ and $\gamma_{R2}$ (purple line cuts), respectively.
These MBSs are represented by the finite and equally amplitudes for
the zero-energy mode at $\omega=0.$ With $J=0,$ ${\cal {A}}_{\gamma_{L2}}$
and ${\cal {A}}_{\gamma_{R1}}$ display a split-peak structure as
a function of $\omega,$ once $\gamma_{L2}$ and $\gamma_{R1}$ builds
the MF dimer (cyan line cuts in Figs.\ref{Fig:Fig.4}(b) and (d)).
For $J\neq0$ and $\omega\neq0$, extra bottom and top arcs rise in
Figs.\ref{Fig:Fig.4}(a), (b) and (d) for ${\cal {A}}_{\gamma_{L1}},{\cal {A}}_{\gamma_{L2}},$
and ${\cal {A}}_{\gamma_{R1}},$ respectively, as a consequence of
the MF trimer formation (green line cuts). Particularly, the amount
of arcs is $2S+1$ due to the quantum spin and it corresponds to half
of the fine structure expected. This issue we shall address later
on. The spill over-like behavior of $\gamma_{L1}$ on $\gamma_{R1}$
can be noted in Figs.\ref{Fig:Fig.4}(a) and (d), where the unbalance
${\cal {A}}_{\gamma_{L1}}(0)<{\cal {A}}_{\gamma_{R1}}(0)$ is notorious
away from $J=0.$ In this manner, we reveal that the zero mode of
the \textit{``Poor M.M.''} continues to spill over from one QD to
another, being induced by $J,$ but with the mode pinned at $\omega=0$
despite the strength of $J.$ Thus, the zero-energy mode of the \textit{``Poor
M.M.''} does not mix with the explicit $2S+1$ fine structure and
characterizes the feature under protection against $J$ coupling.

\begin{figure}[!]
\centering\includegraphics[width=1\columnwidth]{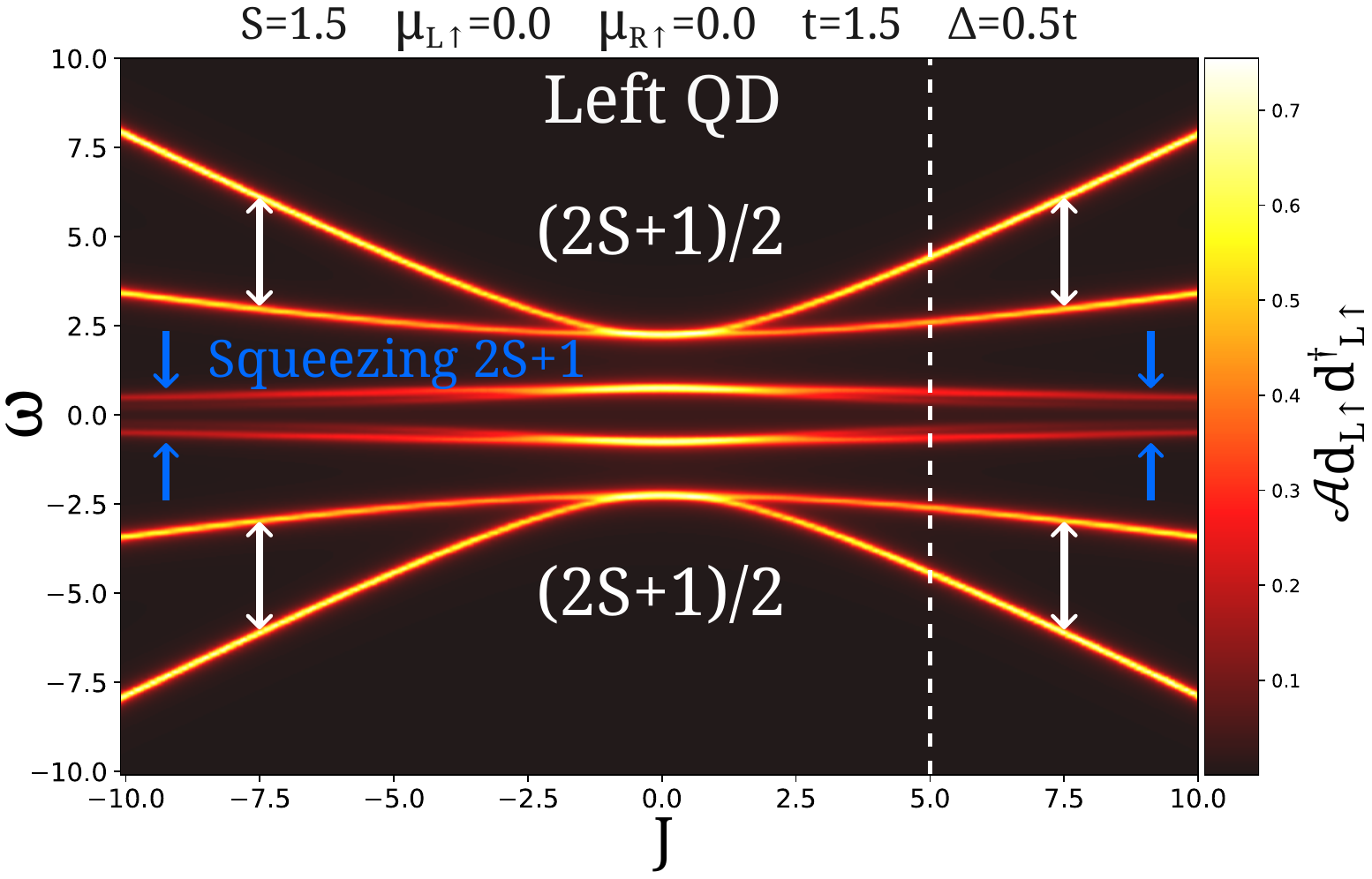} \caption{\label{Fig:Fig.5} Color map of ${\cal {A}}_{d_{L\uparrow}d_{L\uparrow}^{\dagger}}$
off the \textit{\textquotedblleft Majorana chain regime\textquotedblright}
spanned by $\omega$ and $J$ showing the squeezing mechanism of half
of the fine structure towards $\omega=0$ to form the \textit{\textquotedblleft Poor
Man's Majorana\textquotedblright}. As expected, the fine structure
contains the amount of $2\times(2S+1)$ levels, but for $|J|\sim5.$
The inner $(2S+1)$ makes explicit a trend to merge as a zero mode.}
\end{figure}

Back to the fine structure subject, we highlight that the bonding
(bottom arc) and anti-bounding (top arc) states of Figs.\ref{Fig:Fig.2}(c)
and (f) are expected to split into $2S+1$ levels each in the presence
of the quantum spin $S.$ However, such a value is renormalized by
its half $(2S+1)/2,$ as marked by the double arrows in Figs.\ref{Fig:Fig.4}(c)
and (f). It means that the other half is kept squeezed at $\omega=0$
as the zero-energy mode of the \textit{``Poor M.M.'' }and consequently,
it ensures its pinning there, as well as the protection against the
remaining fine structure. From experimental perspective, the unbalance
${\cal {A}}_{d_{\alpha\uparrow}d_{\alpha\uparrow}^{\dagger}}(0)<{\cal {A}}_{d_{\bar{\alpha}\uparrow}d_{\bar{\alpha}\uparrow}^{\dagger}}(0)$
off $J=0$ and absence of mixing characteristic with the explicit
half of the fine structure would emerge as the hallmarks of the \textit{``Poor
M.M.'' }at the QD $\alpha$ exchange coupled to the quantum spin.

In order to elucidate the aforementioned squeezing mechanism of half
of the fine structure responsible to build the \textit{``Poor M.M.''},
in Fig.\ref{Fig:Fig.5} we present ${\cal {A}}_{d_{L\uparrow}d_{L\uparrow}^{\dagger}}$
for the case $\Delta=0.5t,$ which corresponds to a situation off
the \textit{``Majorana chain regime''} . It is worth mentioning
that the exhibition of ${\cal {A}}_{d_{R\uparrow}d_{R\uparrow}^{\dagger}}$
is redundant, once it shares the same features observed in ${\cal {A}}_{d_{L\uparrow}d_{L\uparrow}^{\dagger}}.$
Additionally, we would like to remind that due to the quantum spin
$S,$ the dimer of QDs of Fig.\ref{Fig:Fig.1} is predicted to exhibit
$2\times(2S+1)$ levels for $J\neq0,$ where the number $2$ accounts
for the bounding and anti-bounding states. Therefore, as we can observe
in Fig.\ref{Fig:Fig.5}, this quantity of levels appears as expected,
but resolved just around $|J|\sim5.$ This set is surprising, once
it reveals the trend of the inner fine structure, which delimits precisely
the half $2S+1$ levels from the entire fine structure nearby $\omega=0.$
We call attention that the pattern of this inner portion precedes
the formation of the \textit{``Poor M.M.''}, which is complete in
the \textit{``Majorana chain regime'' }$\Delta=t$ of Fig.\ref{Fig:Fig.4}
with all these $2S+1$ inner levels squeezed at $\omega=0.$ In this
manner, we point out that the inner half-structure is squeezing towards
$\omega=0$ to end-up as the zero-energy mode of the \textit{``Poor
M.M.''} upon approaching $\Delta\rightarrow t.$ Particularly for\textit{
}$\Delta\neq t,$ the anti-crossing point at $J=0$ of the inner levels
arises from the existence of two MF dimers, namely $\gamma_{L2},\gamma_{R1}$
and $\gamma_{L1},\gamma_{R2}.$

For completeness, we clarify that the case of the half-integer $S$
here adopted does not cause loss of generality. In the situation of
an integer spin, the explicit fine structure would have only the mode
$\omega=0$ degenerate with the corresponding squeezed at this point
instead and the phenomenon reported would be still observable.

To summarize, in the Supplemental Material we have prepared an animated
plot for the vertical line cut depicted in Fig.\ref{Fig:Fig.5}, where
the crossover from the regime $\Delta=0,$ with the $2\times(2S+1)$
fine structure, evolves towards the situation $\Delta=t,$ in which
the half $(2S+1),$ then leads to the \textit{``Poor M.M.''} at
$\omega=0.$

\textit{Conclusions.-} We reveal in the system of two superconducting
QDs that the \textit{``Poor Man's Majorana''} exhibits a local protection
against the spin splitting when its host QD is exchange coupled to
a quantum spin. Two key features ensure such a protection. First,
the MBS zero mode consistently remains at zero frequency regardless
of the coupling strength with the spin. Second, the unexpected half
of the fine structure does not eliminate the zero-energy mode. We
attribute such a behavior to the squeezing phenomenon of the other
half as the zero-energy mode of the \textit{``Poor Man's Majorana''}.
It keeps the mode detached from the remaining and observable portion
of the fine structure, making the MBS protected. Our findings pave
the way in highlighting that such a MBS is robust against the generally
supposed unavoidable split by the fine structure from the spin. To
conclude, we understand that these results shift the lack of topological
protection paradigm of the \textit{``Poor Man's Majorana''}, thereby
unveiling new possibilities for this quasiparticle.

\textit{Acknowledgments.-} We thank the Brazilian funding agencies
CNPq (Grants. Nr. 302887/2020-2, 303772/2023-9, 311980/2021-0, and
308695/2021-6), the São Paulo Research Foundation (FAPESP; Grant No.
2023/13467-6), Coordenação de Aperfeiçoamento de Pessoal de Nível
Superior - Brasil (CAPES) -- Finance Code 001 and FAPERJ process
Nr. 210 355/2018. LSR acknowledges the support from the Icelandic
Research Fund (Rannís), Grant No. 239552-051. LSR thanks Unesp for
their hospitality. HS acknowledges the project No. 2022/45/P/ST3/00467
co-funded by the Polish National Science Centre and the European Union
Framework Programme for Research and Innovation Horizon 2020 under
the Marie Sk\l odowska-Curie grant agreement No. 945339.

\bibliographystyle{apsrev4-2}

\end{document}